\documentclass[twocolumns,10pt]{IEEEtran}

 \usepackage{float}
 
\usepackage{amsbsy}
\usepackage{amssymb}
\usepackage{amscd}
\usepackage{amsmath}
  
\usepackage{cite,enumerate}
 
 \usepackage{tikz}

\def\tu{{{t_{\scriptsize\mbox{u}}}}}
\def\ru{{{r_{\scriptsize\mbox{u}}}}}
  
\def\mtp{{{\mbox{mod}\ 2 \pi}}}
  
\newcommand{\arrayantenna}[2]{
\draw[thick,-] (#1,#2) -- (#1+1.0,#2);
\draw[thick,-] (#1,#2) -- (#1,#2+.2);
\draw[thick,-] (#1+.2,#2) -- (#1+.2,#2+.2);
\draw[thick,-] (#1+.4,#2) -- (#1+.4,#2+.2);
\draw[thick,-] (#1+.6,#2) -- (#1+.6,#2+.2);
\draw[thick,-] (#1+.8,#2) -- (#1+.8,#2+.2);
\draw[thick,-] (#1+1.0,#2) -- (#1+1,#2+.2);

\draw[thick,-] (#1.25,#2-.5) -- (#1+.5,#2);
\draw[thick,-] (#1.75,#2-.5) -- (#1+.5,#2);

}

\newcommand{\terminal}[2]
  {\draw[thick] (#1,#2) rectangle (#1-.125,#2-.25);
\draw[thick] (#1,#2) -- (#1,#2+.1);
}

\usepackage[color=black,opacity=1]{background}

\SetBgContents{ \begin{minipage}{35cm}{\copyright 2023 IEEE. Personal use
      of this material is permitted. Permission from IEEE must be obtained
       for all other uses, in any current or future media, including
        reprinting/republishing this material for advertising or promotional purposes,
         creating new collective works, for resale or redistribution to servers or lists, 
         or reuse of any copyrighted component of this work in other works. This paper will appear in the IEEE   Communications Letters, 2023.}\end{minipage}}
\SetBgScale{.5}
\SetBgAngle{0}
\SetBgPosition{current page.south west}
\SetBgHshift{20cm}
\SetBgVshift{2cm}

\title{Phase Calibration of Distributed Antenna Arrays}

\author{Erik G. Larsson$^*$ and Joao Vieira$^\dagger$ \thanks{This work was supported  by ELLIIT, the KAW Foundation, and
 the European Union’s Horizon 2020 research and innovation programme under grant agreement No 101013425 (REINDEER). \\ $^*$Link\"oping University, Dept. of Electrical Engineering (ISY),  581 83 Link\"oping, Sweden. E-mail: \texttt{erik.g.larsson@liu.se}. \\    $^\dagger$Ericsson Research, Mobilv\"agen 12, 223 62 Lund, Sweden. \\ E-mail: \texttt{joao.vieira@ericsson.com}.}}

\begin{document}
 
	\maketitle
	
\begin{abstract}
Antenna arrays can be either reciprocity calibrated (R-calibrated),
which facilitates reciprocity-based beamforming, or fully calibrated
(F-calibrated), which additionally facilitates transmission and
reception in specific physical directions.  We first expose, to
provide context, the fundamental principles of over-the-air R- and
F-calibration of distributed arrays.  We then describe a new method
for calibration of two arrays that are individually F-calibrated, such
that the combined array becomes jointly F-calibrated.
\end{abstract}	
	
\section{Introduction}

Coherent beamforming with antenna arrays is a cornerstone technology
in co-located and distributed (massive) MIMO systems.  In co-located MIMO, the
antennas are located close by one another in an array; in distributed
MIMO, they are spread out geographically, typically in the form of
interconnected antenna panels that together constitute a large array.

For antenna arrays to work efficiently, they must be calibrated in
amplitude and phase --  the phase is the most important.
There are two types of phase calibration: reciprocity (R) calibration, and full (F) calibration.
R-calibration enables reciprocity-based operation, relying on  uplink
pilots for downlink beamforming.  F-calibration is stronger, and
additionally enables the use of geometrically parameterized array models.

We are concerned with calibration of antenna arrays via over-the-air
(OtA) measurements.  OtA calibration is attractive as it circumvents
the use of synchronization cables.  To provide context, and for
completeness, we first give a unified treatment of the basic
principles of R- and F-calibration OtA, using elementary mathematics
and a phasor formalism (Sections~\ref{sec:model}--\ref{sec:calibrone}) -- something missing
from the literature.  We then present our main technical contribution:
a method for OtA F-calibration of two individually F-calibrated arrays
(Section~\ref{sec:calibrtwo}).  Finally, we dispel some misconceptions
that we have encountered when working on array calibration
(Section~\ref{sec:mis}).
 
\section{Preliminaries}\label{sec:model}

\begin{figure}[t!]   
\begin{center}
  \begin{tikzpicture}[xscale=.7,yscale=0.7]
    \arrayantenna{0}{0};
\terminal{0.2}{1};
  \end{tikzpicture}
  \hspace*{2cm}
  \begin{tikzpicture}[xscale=.7,yscale=0.7]
    \arrayantenna{0}{0};
   \begin{scope}[rotate=250,shift={(-1,-1)}]   \arrayantenna{0}{0}; \end{scope}
 \terminal{1}{1};
 
  \end{tikzpicture}
  
\end{center}
\caption{Two instances of an array:  six co-located antennas A$_1$,\ldots,A$_6$ (left);
   twelve antennas A$_1$,\ldots,$A_{12}$  in two distributed panels (right).\label{fig:system}}
   \vspace*{-5mm}
   \end{figure}
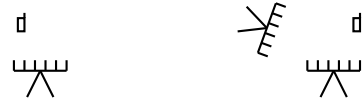

We consider a system with a mobile user and a set of service antennas,
$\{\mbox{A}_i\}$.  For now we make no assumptions on the 
locations of the antennas: they may be co-located, or distributed.  In
either case, we refer to  $\{\mbox{A}_i\}$ collectively as the
\emph{array} (Figure~\ref{fig:system}).

Each antenna $\{\mbox{A}_i\}$ has a transmit and a receive branch; the
same goes for the user.  To each antenna we associate two local
phasors that revolve $f_c$ times per second: one for the transmit
branch and one for the receive branch.  Similarly, to the user we
associate a local transmit phasor and a local receive phasor.
Additionally, we consider a \emph{fictitious}, global, unobservable
phasor that represents an \emph{absolute} reference.  This global
phasor does not exist physically, but serves as a useful abstraction.
Note that the notion of a revolving (rotating) phasor is the standard one used in physics and engineering \cite{mcclellan2017dsp}.

Define:
\begin{itemize}
\item $\tu $, $\ru $: the user's transmit and receive phasor values when the global phasor points to zero.

\item $t_i$, $r_i$: the transmit and receive phasors of A$_i$ when the
  global phasor is zero.\footnote{The subscript
  $(\cdot)_{\scriptsize{\mbox{u}}}$ refers to the user; $(\cdot)_{i}$
  refers to service antenna  A$_i$.}

\item $T_{ij}$: propagation (coupling) delay between A$_i$ and A$_j$; owing to reciprocity, $T_{ij}=T_{ji}$.
 
\item $T_i$: propagation delay between the user and A$_i$. 
\end{itemize}
In general, $\tu
\neq \ru $, $t_i\neq r_i$, and $t_i \neq t_j$ because of differences
in the circuitry, and because of oscillator drifts.

We consider  passband signals with carrier
frequency $f_c$. Hence, we can equivalently refer to time as phase:
phase equals ($2\pi f_c \times$time) $\mtp$.  A ``signal'' at an
antenna means the voltage across the antenna terminal.  When saying that
a signal is received (transmitted) ``at local [or global] phase
$\phi$'', we mean that the signal undergoes a positive zero-crossing
when the corresponding local [or global] phasor points to $\phi$.  All
quantities are in radians, and all phasor values and phase
measurements are defined $\mtp$.  Our focus will be  on identifiability
questions, thus we neglect measurement noise. 

Throughout, \emph{propagation delay} between a transmit and receive
antenna means the phase lag between the corresponding signals.  This
delay includes the effect of the antenna radiation patterns, which are
completely reciprocal.  If antennas are closely located, the delay is
mainly determined by the mutual coupling.  In this paper,
\emph{calibration} means determining relevant functions of
$\{t_i,r_i\}$ to facilitate phase-synchronous transmission or
reception, or both.  When calibrating multiple arrays together we will
use the term ``phase alignment''.

Rather than working with $\{t_i,r_i\}$, some related literature works
with ``calibration coefficients'' $\{\gamma_{t_i}, \gamma_{r_i}\}$
defined in complex baseband \cite{RBPCMMBP:14:TWC,vieralarsson_pimrc}.
All results in this paper can be re-derived using that formalism
instead, by setting $\gamma_{r_i}=e^{j r_i}$ and $\gamma_{t_i}=e^{-j
  t_i}$. (Note the minus sign: a positive shift of $t_i$ has the
opposite effect on the signal phase as a positive shift of $r_i$,
since the signals travel in opposite directions.) This means, for example, that the statement
``$r_i=t_i$'' is equivalent to ``$\gamma_{r_i}\gamma_{t_i}=1$''.  The
advantage of the phasor formalism is that it connects more naturally
to fundamental physics, and yields simple linear equations.
 
\subsection{Fully Calibrated Array}

If the differences $\{r_i-r_j\}$ are known for all pairs $(i,j)$, we
say that the array is \emph{fully receive-calibrated} (FR-calibrated).
The circuitry can then post-compensate the received signals such that
we can without loss of generality assume that $r_i=c$ for all antennas
A$_i$ and some indeterminable constant $c$.  This ensures that an
observed phase-shift between uplink signals received at two antennas
can be mapped onto a specific physical direction, 
facilitating  angle-of-arrival estimation and positioning
based on fingerprinting. Note that $\tu $ does not have to be known,
as it  rotates the signal phase at all antennas.

Analogously, if $\{t_i-t_j\}$ are known for all $(i,j)$ we have a
\emph{fully transmit-calibrated} (FT-calibrated) array.
FT-calibration is required for transmission in specific physical
directions, for example, for grid-of-beams beamforming.

If $\{r_i-r_j\}$ and $\{t_i-t_j\}$ are known, then the array is both
FR- and FT-calibrated.  If, in addition, all differences $\{r_i-t_i\}$
are known, we say that we have a \emph{fully calibrated}
(F-calibrated) array.  
For an F-calibrated array, we can assume, after appropriate post- and
pre-compensation of received and transmitted signals, that $r_i=t_i=c$
for all $i$ and some indeterminable constant $c$. Namely, let $c=r_1$
(which is unknown); compensate the phase of A$_1$'s transmitter chain
such that after compensation, $t_1=r_1$ (possible since
$t_1-r_1$ is known); compensate the phase of A$_2$'s receive chain
such that $r_2=r_1$ (possible since $r_2-r_1$ is
known); and so forth; eventually, $r_i=t_i=c$ for all $i$.

\subsection{Reciprocity Calibrated Array}

If $\{t_j-t_i+r_j-r_i\}$ are known for all antenna pairs $(i,j)$, we
say that the array is \emph{reciprocity calibrated}
(R-calibrated). This is the exact condition required for
reciprocity-based beamforming to work on downlink. To see why, it is
sufficient to consider two service antennas, say, A$_1$ and A$_2$.

First, on uplink, the user transmits a pilot.  Suppose that this pilot
is transmitted at local phase (at the user) $0$; when the local phasor
is $0$, the global phasor points to $-\tu $.  A$_1$ observes the signal
at global phase $T_1-\tu $, which is (at A$_1$) local phase
$r_1+T_1-\tu $.  Similarly, A$_2$ receives the signal at local
phase $r_2+T_2-\tu $.

Next, on downlink, A$_1$ and A$_2$ apply ``conjugate beamforming'',
shifting the phase of the transmitted signal by the respective negated
observed pilot phase.  This means that A$_1$ transmits at local phase
$-(r_1+T_1-\tu )$, which is global phase $-t_1-r_1-T_1+\tu $.  The
user observes the signal at global phase $-t_1-r_1+\tu $, which is
local phase $-t_1-r_1+\tu +\ru $.  Similarly, A$_2$ transmits at
global phase $-t_2-r_2-T_2+\tu $; this signal is observed by the user
at local phase $-t_2-r_2+\tu +\ru $.

For the signals from A$_1$ and A$_2$ to add upp constructively, A$_1$
and A$_2$ need to pre-compensate the phases of their  transmitted signals by $t_1+r_1+c$ and
$t_2+r_2+c$, for some arbitrary $c$.  So what needs to
be known is the difference between $t_1+r_1+c$ and $t_2+r_2+c$, that is,
$t_1-t_2+r_1-r_2$.  With more antennas, we need to know
$\{t_1-t_i+r_1-r_i\}$ for all $i$, or equivalently,
$\{t_j-t_i+r_j-r_i\}$ for all $(i,j)$. Note how the conjugate
beamforming eliminates $T_1$ and $T_2$.  The coherently combined signal received at the
user is phase-rotated by $\tu+\ru+c$; this residual phase shift can be
eliminated by using a demodulation pilot or by blind decoding.

Note that FR-plus-FT-calibration (or F-calibration) implies
R-calibration: if $\{r_i-r_j\}$ and $\{t_i-t_j\}$ are known for all
$(i,j)$, then $\{t_j-t_i+r_j-r_i\}$ can be computed for all $(i,j)$.
The converse, however, does not hold.
 
\section{A Unified Description of  OtA Calibration}\label{sec:calibrone}

\subsection{F-calibration Methodology}

Methods for F-calibration OtA date back quite far
\cite{aumann1989phased,vieira2016receive}.  These methods require at least three
antennas, and the propagation (coupling) delay between these antennas,
$\{T_{ij}\}$, must be known.  Obtaining $\{T_{ij}\}$ is feasible for a
co-located array, although it requires electromagnetic simulations or
measurements in an anechoic chamber.
  
The basic principle of OtA F-calibration is to perform bi-directional
measurements between pairs of antennas.  First, A$_1$ transmits at
local phase $0$, which is global phase $-t_1$. A$_2$ receives at
global phase $-t_1+T_{12}$, which is local phase $r_2-t_1+T_{12}$;
after subtracting the known $T_{12}$, we obtain the measurement
$r_2-t_1$. Continuing for all three pairs, in both directions, we
obtain  six measurements:
\begin{align}
 d_{12} & = r_2 - t_1 ,  &  d_{21} & = r_1 - t_2 , &  d_{13} & = r_3 - t_1 \label{eq:bidirec1} \\
  d_{31} & = r_1 - t_3 ,  &
 d_{32} & = r_2 - t_3 , & d_{23} &  = r_3 - t_2 .\label{eq:bidirec1b}
 \end{align}
Equations (\ref{eq:bidirec1})--(\ref{eq:bidirec1b}) are easily rearranged to obtain 
\begin{align}
r_1-t_1 & = d_{21}+d_{13}-d_{23} , \\
r_1-r_2 & = d_{31}-d_{32} , \\
t_1-t_2 & = d_{23}-d_{13},
\end{align}
and so forth.
This way, we can determine  $\{r_i-r_j\}$, $\{t_i-t_j\}$, and  $\{r_i-t_i\}$, as required for the array to be F-calibrated.

The above equations can also be found in
\cite{interdonato2019ubiquitous}, although
\cite{interdonato2019ubiquitous} failed to point out the crucial fact
that $\{T_{ij}\}$ must be known to solve for $\{r_i-r_j\}$,
$\{t_i-t_j\}$, and $\{r_i-t_i\}$ -- something that seems infeasible
with distributed antennas.

\subsection{R-calibration Methodology}

R-calibration can be performed using pairwise OtA measurements between
the antennas, without knowing $\{T_{ij}\}$, both for co-located and
distributed arrays
\cite{Kaltenberger,vieira2017reciprocity,Zetterberg:2011:EIT:1928509.1972710,shepard2012argos,RBPCMMBP:14:TWC,vieralarsson_pimrc,chen2017distributed}.
It is the fortunate fact that $\{T_{ij}\}$ need not be known, that
makes reciprocity-based operation possible in distributed MIMO.

To explain the principle it is sufficient to consider two antennas, say,
A$_1$ and A$_2$.  A bi-directional measurement
is performed in  the same way as for F-calibration, but now, $T_{12}$ cannot be subtracted.
This gives the two observations,
\begin{align}
  d_{12} & = r_2-t_1+T_{12} ,  &  d_{21} & = r_1-t_2+T_{12}   .   \label{eq:R1}
 \end{align}
Subtracting $d_{12}$ from $d_{21}$  yields the required quantity
\begin{align}
d_{21}-d_{12}  = t_1-t_2+r_1-r_2 .
\end{align}
For more than two antennas, one continues  by
bi-directional measurements between A$_1$ and A$_i$ for all $i$, or
between other, appropriately selected, pairs.  From these
measurements, $\{t_j-t_i+r_j-r_i\}$ can be computed for all $(i,j)$.
 
\section{Phase-Aligning Two R- or F-Calibrated Arrays}\label{sec:calibrtwo}
 
\begin{figure}[t!]   
\begin{center}
  \begin{tikzpicture}[xscale=.7,yscale=0.7,rotate=15]
  \begin{scope}[shift={(2,0)}]     \begin{scope}[rotate=30]   \arrayantenna{0}{0}; \end{scope} \end{scope}
  \begin{scope}[shift={(0.3,0)}]     \begin{scope}[rotate=0]   \arrayantenna{0}{0}; \end{scope} \end{scope}
  \begin{scope}[shift={(-2,2)}]     \begin{scope}[rotate=250]   \arrayantenna{0}{0}; \end{scope} \end{scope}

    \draw[dotted,thick,rotate around={20:(1.5,0)},black](1.5,0) ellipse (2 and 1 );
    \draw[dotted,thick,rotate around={20:(-2.2,1.5)},black](-2.2,1.5) ellipse (1.5 and 1 );
  \node at (-4,2) {A};
  \node at (3.8,0) {B};
  \node at (4.6,0.9) {AB};
    \draw[dashed,thick,rotate around={-15:(1.5,0)},black](-0.2,0.2) ellipse (5 and 1.85);
  \end{tikzpicture}
  
\end{center}
\caption{Two arrays A and B, collectively constituting the array AB.\label{fig:AB}}
   \end{figure}
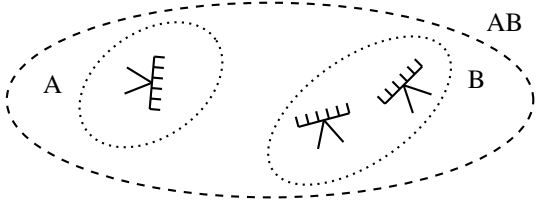

We now consider the following question: Given \emph{two arrays}
(that may themselves be co-located or distributed), of which each one is either
F-calibrated or R-calibrated, \emph{what is required to calibrate the two
arrays relative to one another?}  We call the two arrays A and B, and
their respective antennas $\{\mbox{A}_i\}$ and $\{\mbox{B}_i\}$; we
write AB for the array collectively comprised by $\{\mbox{A}_i\}$ and
$\{\mbox{B}_i\}$ (Figure~\ref{fig:AB}).  This problem is of specific  interest
when A and B are antenna panels (``access points'') in distributed MIMO, in
which case the calibration task is sometimes called \emph{phase alignment}.
Nothing precludes A or B from being a mobile user, although that may
not be a typical scenario. In what follows, we assume that there is a central entity that
can co-process measurements from both A and B.

If A and B are separated, then $\{T_{A_iB_j}\}$ cannot reasonably be known a priori.
Yet, as we will see in Section~\ref{sec:AAAB},  F-calibration of the joint array AB is possible.

\subsection{Aligning Two R-Calibrated Arrays}\label{sec:RAAR}

First consider the case when  A and B are individually R-calibrated.
Clearly, AB cannot then be made  F-calibrated.
But A and B can be  phase-aligned such that AB  becomes R-calibrated.
This only takes a simple bi-directional measurement between any of $\{\mbox{A}_i\}$ and any of $\{\mbox{B}_j\}$:
\begin{align}
 d_{A_i B_j}  & = r_{B_j}-t_{A_i}+T_{A_i B_j}  , \label{eq:twoRbidia}  \\
 d_{B_j A_i}   & = r_{A_i}-t_{B_j}+T_{A_i B_j} \label{eq:twoRbidib}.
\end{align}
Say we use A$_1$ and B$_1$; this gives the measurements $ d_{A_1 B_1}
$ and $ d_{B_1 A_1}$.  Subtraction of $d_{A_1 B_1}$ from $d_{B_1 A_1}
$ gives $t_{A_1}-t_{B_1}+r_{A_1}-r_{B_1}$.  Once
$t_{A_1}-t_{B_1}+r_{A_1}-r_{B_1}$ is known, along with
$\{t_{A_i}-t_{A_j}+r_{A_i}-r_{A_j}\}$ and
$\{t_{B_i}-t_{B_j}+r_{B_i}-r_{B_j}\}$ (which are known by assumption),
$\{t_{A_i}-t_{B_j}+r_{A_i}-r_{B_j}\}$ can be computed for any $(i,j)$
by pairwise addition and subtraction.  Alternatively, more
combinations of antenna pairs (A$_i$,B$_j$) can be measured.
Beamforming can be beneficially exploited to improve the SNR if the
arrays are distant from one another, for example using a protocol similar to that in \cite{ganesan2021beamsync}.

A useful fact, though unrelated to reciprocity-based beamforming, is
the following. Once AB is R-calibrated, then after appropriate
compensation, the ``channels'' from $\mbox{A}_i$ to $\mbox{B}_j$ and
from $\mbox{B}_j$ to $\mbox{A}_i$ are equal.  Specifically, consider
the two measurements between $\mbox{A}_i$ and $\mbox{B}_j$ in
(\ref{eq:twoRbidia})--(\ref{eq:twoRbidib}). They differ precisely by
$t_{A_i}-t_{B_j}+r_{A_i}-r_{B_j}$, which is known, and hence can be
compensated for.

\subsection{Aligning Two F-Calibrated Arrays}

Next, consider the case when A and B are (individually) F-calibrated.
We can assume, after appropriate pre- and post-compensation of the
transmitted and received signals, that $t_{A_i}=r_{A_i}=c_A$ and
$t_{B_i}=r_{B_i}=c_B$ for all $i$ and some indeterminable constants
$c_A$ and $c_B$.

To illustrate the methodology, we pick an arbitrary antenna from A and an arbitrary antenna from B,
enabling us to drop the index $i$.  Anything said
in what follows, if desired, can be repeated for any other pair (A$_i$,B$_j$) of
antennas, and followed by appropriate averaging over antenna pairs to
suppress measurement noise, and consequently obtaining a processing gain that improves the
accuracy.
\\

\subsubsection{R-Calibration of the Joint Array AB}
\label{sec:RAAB}

Since F-calibration is stronger than R-calibration, it comes as no
surprise that one can  phase-align A
and B, such that AB becomes R-calibrated.  
The goal is to obtain,
\begin{align}\label{eq:cAcB}
t_{A}-t_{B}+r_{A}-r_{B}  = 2(c_A-c_B),
\end{align}
without knowledge of  $\{T_{A_iB_j}\}$. This is easily achieved by a 
bi-directional measurement between A and B:  
\begin{align}
d_{AB} & =r_B - t_A +T_{AB}    = c_B - c_A + T_{AB}, \label{eq:singmeas1} \\
d_{BA}& =r_A - t_B +T_{AB} = c_A - c_B + T_{AB}. \label{eq:singmeas2}
\end{align}
Subtracting (\ref{eq:singmeas1}) from (\ref{eq:singmeas2})   yields  the sought-after quantity required for AB to be R-calibrated (see (\ref{eq:cAcB})):
\begin{equation}\label{eq:combmeas}
d_{BA}-d_{AB}  = 2(c_A-c_B).
\end{equation}

Note that had $T_{AB}$ been known (which is unrealistic in practice),
then $c_A-c_B$ could have been immediately obtained from a
\emph{single} measurement (\ref{eq:singmeas1}) or (\ref{eq:singmeas2})
alone. \\

\subsubsection{F-Calibration of the Joint Array AB}\label{sec:AAAB}

Now consider the more intricate case when A and B are individually
F-calibrated and we want to phase-align A and B such that AB becomes
F-calibrated.  
We must then, \emph{without knowing $T_{AB}$}, obtain $c_A-c_B$, rather than $2(c_A-c_B)$.  This is
not possible from a bi-directional measurement as in
(\ref{eq:singmeas1})--(\ref{eq:singmeas2}), since dividing
(\ref{eq:combmeas}) by $2$ yields an ambiguity of mod $\pi$.

Our proposed resolution of this problem is to use   measurements
with two signals, say two sinusoids with frequencies $f$ and $f'$,
where $f'<f$.  (In an OFDM system, $f$ and $f'$ could correspond to
different subcarriers.)  The frequencies $f$ and $f'$ are associated
with different propagation delays (phase lags), say $T_{AB}$ and
$T'_{AB}$ (in radians).  Specifically,
\begin{align}
  T_{AB}'=\frac{f'}{f}T_{AB}.
  \end{align}
Since $f'/f<1$, a higher frequency means a shorter wavelength, which
means a larger propagation delay when expressed in radians.  In the
presence of multipath, $f-f'$ must be less than the channel coherence
bandwidth (reciprocal delay spread).

Taking a  bi-directional measurement at frequency $f$ (similar to in Section~\ref{sec:RAAB}), and a uni-directional measurement at frequency
$f'$, yields three phase measurements ($\mtp$),
\begin{align}
d_{AB} &=c_B-c_A+T_{AB} , \label{eq:bidi1a} \\ 
d_{BA} &=c_A-c_B+T_{AB}  , \label{eq:bidi1b} \\
d'_{BA} &=c_A-c_B+T'_{AB}    \label{eq:bidi1d}
\end{align}
where $(d_{AB}, d_{BA})$ are obtained by measurements on the
$f$-signal, and $d'_{BA}$ is obtained from the $f'$-signal.

There are three unknowns: $c_A-c_B$, $T_{AB}$ and $T'_{AB}$, where the first is of interest, and the latter two are  nuisance parameters (which are proportionally related).
By   subtracting  (\ref{eq:bidi1a}) from (\ref{eq:bidi1b}),  $T_{AB}$ disappears  and we conclude that $c_A-c_B$ is one of the following two values  ($\mtp$),
\begin{align}
\widehat{(c_A-c_B) }_i  &= (d_{BA}-d_{AB} )/2 ,  \label{eq:bidi2a} \\
\widehat{(c_A-c_B) }_{ii}   & =  \pi + (d_{BA}-d_{AB} )/2  . \label{eq:bidi2b}
\end{align}

To determine which one of $\widehat {(c_A-c_B) }_i $ and $\widehat {(c_A-c_B) }_{ii} $  in (\ref{eq:bidi2a})--(\ref{eq:bidi2b}) is the correct one, we proceed as follows. 
By adding   (\ref{eq:bidi1a}) and (\ref{eq:bidi1b}),  we infer that $T_{AB}$ is one of the two values    ($\mtp$),
\begin{align}
T_{AB}^i & = (d_{AB} + d_{BA})/2 , \quad \mbox{or} \label{eq:bidi3a} \\
T_{AB}^{ii} & = \pi + (d_{AB} + d_{BA})/2 .\label{eq:bidi3b}
\end{align}
But from (\ref{eq:bidi1b}) and (\ref{eq:bidi1d}), we  concurrently deduce that 
\begin{align}
(1-f'/f )T_{AB}=T_{AB}-T'_{AB} =d_{BA}-d'_{BA}
\end{align}
($\mtp$), from which it follows that
\begin{align}
T_{AB}=\hat{T}_{AB} \ \mbox{mod}\  \frac{2\pi}{  1-f'/f}, \label{eq:bidi4}
\end{align}
where we defined
\begin{align}
\hat{T}_{AB}= \frac{d_{BA}-d'_{BA}}{1-f'/f}.\label{eq:bidi5}
\end{align}
We can then check which of $T_{AB}^i+m2\pi$ and $T_{AB}^{ii}+m2\pi$  in (\ref{eq:bidi3a})--(\ref{eq:bidi3b})  matches most closely with  $\hat{T}_{AB}+n2\pi/ (1-f'/f )$, for some combination of integers $m$ and $n$. (Absent measurement noise, one of them will match exactly.) 
If the match is best for $T_{AB}^i$  then we conclude that,
\begin{align}
T_{AB}=T_{AB}^i\ \mtp
\end{align}
(hereafter referred to as {case (i)}); otherwise
\begin{align}
  T_{AB}=T_{AB}^{ii}\ \mtp
\end{align}
(referred to as {case (ii)}). 
In case (i), insertion of $T_{AB}$ into (\ref{eq:bidi1a})  gives that ($\mtp$)
\begin{align}
c_A - c_B = T_{AB}-d_{AB} = (d_{BA}-d_{AB} )/2  , 
\end{align}
i.e., $c_A-c_B = \widehat{ (c_A - c_B) }_i $.
In case (ii), (\ref{eq:bidi1a})  instead gives that 
\begin{align}
c_A - c_B = T_{AB}-d_{AB}  = \pi + (d_{BA}-d_{AB} )/2,
\end{align}
that is, $c_A-c_B = \widehat{ (c_A - c_B) }_{ii}$.

The physical interpretation of the algorithm is that we estimate, mod $2\pi/ (1-f'/f )$ radians, the propagation delay between A and B, using a probing signal with bandwidth $f-f'$. 
The  estimated distance is then used as side information when resolving the $\pi$-ambiguity in the phase alignment.

\begin{figure}[t!]
  \centerline{\ \scalebox{0.93}{\input{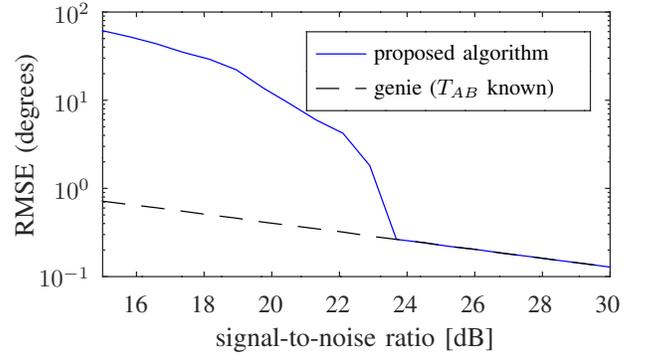}}}
\caption{Simulation example of phase-alignment of two individually
  F-calibrated arrays A and B, such that AB becomes F-calibrated.\label{fig1}}
\vspace*{-5mm}
\end{figure}

The objective here has been to demonstrate how the $\pi$-ambiguity can be
resolved, rather than to describe an algorithm that is optimal in the
presence of  measurement noise with some particular distribution.  Variations and improvements are
possible, and useful if noise is non-negligible. For example, one can obtain a fourth measurement,
making the measurement at frequency $f'$ bi-directional as well:
\begin{align}
d'_{AB} &=c_B-c_A+T'_{AB}.  \label{eq:bidi1c}    
  \end{align}
Then, estimates similar to
(\ref{eq:bidi2a})--(\ref{eq:bidi2b}), expressed as a function of
$(d'_{AB},d'_{BA})$, can be formed from
(\ref{eq:bidi1d}) and (\ref{eq:bidi1c}), instead.  Averages of estimates
formed from (\ref{eq:bidi1a})--(\ref{eq:bidi1b}), and from
(\ref{eq:bidi1d}) and (\ref{eq:bidi1c}) can be formed.  Also, instead of
forming $T_{AB}^i $ and $T_{AB}^{ii}$ from
(\ref{eq:bidi1a})--(\ref{eq:bidi1b}), one can use
(\ref{eq:bidi1d}) and (\ref{eq:bidi1c}) instead, or, use an average of
the two estimates formed this way.  Furthermore, rather than obtaining
$ \hat{T}_{AB}$ in (\ref{eq:bidi5}) from (\ref{eq:bidi1b}) and
(\ref{eq:bidi1d}), one can obtain it from (\ref{eq:bidi1a}) and
(\ref{eq:bidi1c}), or, from the average of two such
estimates. Formulation as a regression problem is also possible.
We have to leave the precise formulation of statistically optimal estimators as future work.

To exemplify the principle, Figure~\ref{fig1} shows a simulation of
the estimation performance for the proposed algorithm, compared
to a genie baseline that knows $T_{AB}$.  We took $f=2$ GHz, $f'=f-50$
MHz, and somewhat arbitrarily, a distance from A to B of $50$
wavelengths, and selected $c_A$ and $c_B$ randomly.  To obtain a
processing gain, 100 samples were coherently averaged for each phase
measurement.  A closest-neighbor fit was used to find the integer pair
$(m,n)$ that defines the best match between $\hat{T}_{AB}+n2\pi/
(1-f'/f )$ and one of the values $\{T_{AB}^i+m2\pi,T_{AB}^{ii}+m2\pi\}$.  Due to
the non-linearity of the ambiguity resolution operation, a threshold
effect is observed. Once the SNR is high enough, the algorithm's
performance meets the genie bound, and improves $\sqrt{10}$ times in RMSE per 10 dB  increase in SNR -- as expected, since once the $\pi$-ambiguity is resolved with high enough probability, the estimate is essentially
a linear function of the noisy observations.

We end by pointing out that the propagation distance must be the same at $f$ and $f'$. 
In practice, one would have to operate in or near line-of-sight conditions, or, if A and B have multiple antennas,  apply beamforming  to ensure that no distant multipath is present. 
Similarly, corrections may have to be applied to compensate for phase variations in the antenna frequency response.

 \section{Misconceptions} \label{sec:mis}

We discuss some misconceptions that we have encountered.

\begin{enumerate}[1.]
\item 
\emph{Myth:} Two individually calibrated arrays A and B can be used together for joint reciprocity-based beamforming. \\
\emph{Reality:}  A bi-directional, OtA measurement is required to phase-align the arrays, 
irrespective of whether the arrays are (individually) R- or F-calibrated.

\item\emph{Myth:}  R-calibration is sufficient to perform  grid-of-beams beamforming with fully digital arrays. \\
\emph{Reality:}  F-calibration is required for this.

\item \emph{Myth:} For reciprocity-based beamforming, A$_i$ should pre-compensate the phase with $t_i-r_i$. 
(For example, \cite{interdonato2019ubiquitous}   states that ``The difference in time reference between the transmitter and receiver in a given AP represents the reciprocity calibration error.'') \\
\emph{Reality:}  A$_i$ should pre-compensate with $t_i+r_i+c$ (for some arbitrary but common-for-all-A$_i$ $c$).  

\item \emph{Myth:} A drift in the oscillator that drives the radio-frequency mixer of A$_i$ does not require 
 reciprocity re-calibration. 
 (For example, \cite{bjornson2015distributed} assumes this.) \\
  \emph{Reality}:  Such a shift affects  $t_i$ and $r_i$ equally, so $t_i+r_i-(t_j+r_j)$,  which must be known for reciprocity, changes.
  Hence, an oscillator drift calls for re-calibration. 
  This misconception, and its prevalence, were also pointed out in \cite{nissel2022correctly}.

\item \emph{Myth:} 
Channel aging (physically moving A$_i$)   has the same effect on the signal phases as a shift in the sampling clock (oscillator phase), of say, $\phi$.   \\
\emph{Reality:}  
Channel aging has the same effect on measurements as   shifting $t_i$ by $+\phi$ and $r_i$ by $-\phi$.
But an oscillator drift is equivalent to  shifting $t_i$ by $+\phi$ and $r_i$ by $+\phi$.
 
\item \emph{Myth:} Beamforming to a user on downlink with an
  R-calibrated array requires demodulation pilots.  \\ \emph{Reality:}
  Demodulation pilots allow for simpler implementation, but they are
  fundamentally unnecessary.  Unless the user is \emph{jointly}
  R-calibrated with the array, the signal received at the user will be
  randomly phase-rotated.  This phase rotation (along with a partially
  unknown amplitude gain) can be estimated   using a demodulation
  pilot,   or with  a joint equalization and decoding algorithm
  \cite{tuchler2011turbo}, or even blindly.  \\
  
Note also that nothing, except implementation complexity, prevents
\emph{joint} R-calibration of the user and the array: take the array to be A
and the user to be B in Section~\ref{sec:RAAR}. In this case, the
phase rotation at the user vanishes, and the remaining unknown
amplitude gain can be estimated blindly through simple methods
\cite{ngo2017no}.

\end{enumerate}

\section{Concluding Remarks}

We considered antenna arrays, in their most general meaning: an array
can be co-located or distributed, and have an arbitrary number of
antennas. We  showed that the concepts of F- and R-calibration OtA can be
rigorously explained using a phasor formalism in which all
relations are described through simple, linear equation systems.  We then
used this framework to develop a method for joint OtA F-calibration of two
arrays A and B that have been previously individually F-calibrated.
Finally, we discussed a number of misconceptions related to array
calibration.

\end{document}